\documentclass[aps]{revtex4}
\usepackage{graphicx,psfrag}
\def\bc{\begin{center}}
\def\ec{\end{center}}

\def\beq{\begin{equation}}
\def\eeq{\end{equation}}

\def\br{{\bf r}}

\def\bk{{\bf k}}

\def\eps{\varepsilon}

\begin{document}

\title{
Lattice symmetries, spectral topology and 
opto-electronic properties of graphene-like materials 
}

\author{K. Ziegler and A. Sinner}
\affiliation{Institut f\"ur Physik, Universit\"at Augsburg\\
D-86135 Augsburg, Germany\\
%(wla short.tex)
}
\date{\today}

\begin{abstract}
The topology of the band structure, which is determined by the lattice symmetries, 
has a strong influence on the transport properties. 
Here we consider an anisotropic honeycomb lattice and study the effect
of a continuously deformed band structure on the optical conductivity and on diffusion due
to quantum fluctuations. In contrast to the behavior at an isotropic node we find 
super- and subdiffusion for the anisotropic node.
The spectral saddle points create van Hove singularities in the optical
conductivity, which could be used to characterize the spectral properties experimentally.
% and discuss several methods for creating a global anisotropy.
\end{abstract}
%\pacs{05.60.Gg, 66.30.Fq, 05.40.-a}

\maketitle

\section{Introduction}
\label{sect:introduction}

Graphene, a two-dimensional sheet of carbon atoms which form a honeycomb lattice
has exceptional opto-electronic properties \cite{novoselov05}. The latter are related
to the bandstructure of this material, consisting of two bands with two Dirac nodes 
\cite{castro09,abergel10}. The existence and the positions of these nodes are a consequence of global 
symmetries of the lattice. Local breaking of the symmetries, for instance, by impurities
or other types of local disorder does not affect the robust opto-electronic properties,
as long as the symmetries are globally preserved. This is the case when the distribution
of the local disorder obeys the global symmetries. The situation changes, though, when
the symmetries are globally broken. A typical example is breaking of the sublattice 
symmetry of the honeycomb lattice when the carbon atoms are replaced by atoms with 
different mass, such that the atomic mass is larger on one sublattice. This leads to
an opening of the Dirac nodes by creating a gap between the two bands. A realization of
this effect is hexagonal Boron Nitride, which is characterized by a gap of 5 eV
\cite{watanabe04}. Another global symmetry of the honeycomb lattice is its isotropy.
Breaking this symmetry by changing the bonds between neighboring sites of the atomic
lattice in one direction affects the positions of the Dirac nodes. For special values
of the anisotropic bonds the Dirac nodes can even brought to the the same position
with only one degenerate node between the two bands. This effect was proposed 
and discussed in a series of papers by Montambaux et al. 
\cite{montambaux09,montambaux09a,delplace10,adroguer16},
and as a Lifshitz transition it was also discussed recently in the context of the 
Kitaev model \cite{faye14}.
In particular, the spectral properties and the DC conductivity become very
anisotropic in the case of the doubly degenerate Dirac node \cite{adroguer16}.
This is quite remarkable in the light of transport properties in graphene-like materials,
which are already exceptional near the Dirac nodes in isotropic graphene. We briefly
summarize the DC transport properties in undoped isotropic graphene. 

Diffusion, the origin of conducting behavior in conventional metals with finite conductivity, 
is a result of random impurity scattering \cite{thouless74}. In the absence of the latter
the diffusion coefficient would diverge and we would experience ballistic transport. 
This simple picture is not valid at spectral degeneracies, though. For instance, 
at a Dirac (Weyl) node or at a node with parabolic spectrum there is diffusion due to quantum 
fluctuations between the upper and the lower band, %\cite{}, 
even in the absence of random impurity 
scattering (cf. Appendix \ref{sect:isotropic}). In this context it should be noted that the
Fermi Golden Rule gives \cite{adroguer16}
\beq
\sigma_{\mu\mu}=\frac{e^2\hbar}{\pi\gamma}v_F^2
\ ,
\label{golden_rule}
\eeq
where $\gamma$ is the strength of the disorder fluctuations and $v_F$ is the Fermi velocity.
It does not reproduce the constant conductivity at the node in %which survives 
the pure limit $\gamma\to0$
\beq
\sigma_{\mu\mu}=\frac{e^2}{\pi h}
\ ,
\label{dirac00}
\eeq
which was experimentally confirmed for graphene \cite{novoselov05,chen08,chen09} (up to the factor $1/\pi$).
The finiteness of the conductivity at the node of the pure system can be attributed to quantum fluctuations between the 
two bands. The latter are not taken into account in the Fermi Golden Rule (\ref{golden_rule}), whereas their
inclusion leads to a finite conductivity at the Dirac node even in the absence of any impurity scattering.
This result reflects a more delicate transport behavior at the spectral node, whereas away from
this node conventional approximations, such as Fermi Golden Rule (\ref{golden_rule}), are valid.

%A new twist:
The result (\ref{dirac00}) of an isotropic nodal spectrum is in stark contrast to the recently found behavior 
for an anisotropic spectrum in the vicinity of a Dirac node.
Adroguer et al. \cite{adroguer16} found a remarkable result for the
conductivity in the presence of random scattering, namely a strongly anisotropic transport behavior,
where $\sigma_{xx}(E_F)$ vanishes linearly with $E_F\to0$, whereas $\sigma_{yy}(E_F)$ stays nonzero even for $E_F\to0$.
We will return to the details of this result in Sect. \ref{sect:model}.

The paper is organized as follows. In Sect. \ref{sect:model} the tight-binding model of an anisotropic
honeycomb lattice is briefly considered, including its relation to a Dirac-like low-energy Hamiltonian.
For this Hamiltonian the optical conductivities (Sect. \ref{sect:opt_cond}) and the diffusion
coefficients (Sect. \ref{sect:diffusion}) are calculated. These results and their relations to 
topological transitions are discussed in Sect. \ref{sect:discussion}.  
\begin{figure}[t]
\begin{center}
\includegraphics[width=4cm,height=4.5cm]{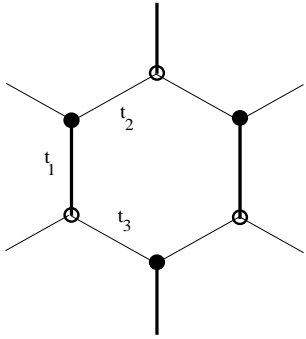}
\caption{\label{fig:hexagon}
%\col
%(Color online) 
A hexagon of the honeycomb lattice with hopping parameters $t_1$, $t_2$ and $t_3$.
The thickness of the bonds refer to the hopping rate. Full (empty) circles indicate the two triangular
sublattices.
}
\end{center}
\end{figure}

\section{Model of merging Dirac nodes}
\label{sect:model}

\begin{figure}[t]
\begin{center}
\includegraphics[width=9cm]{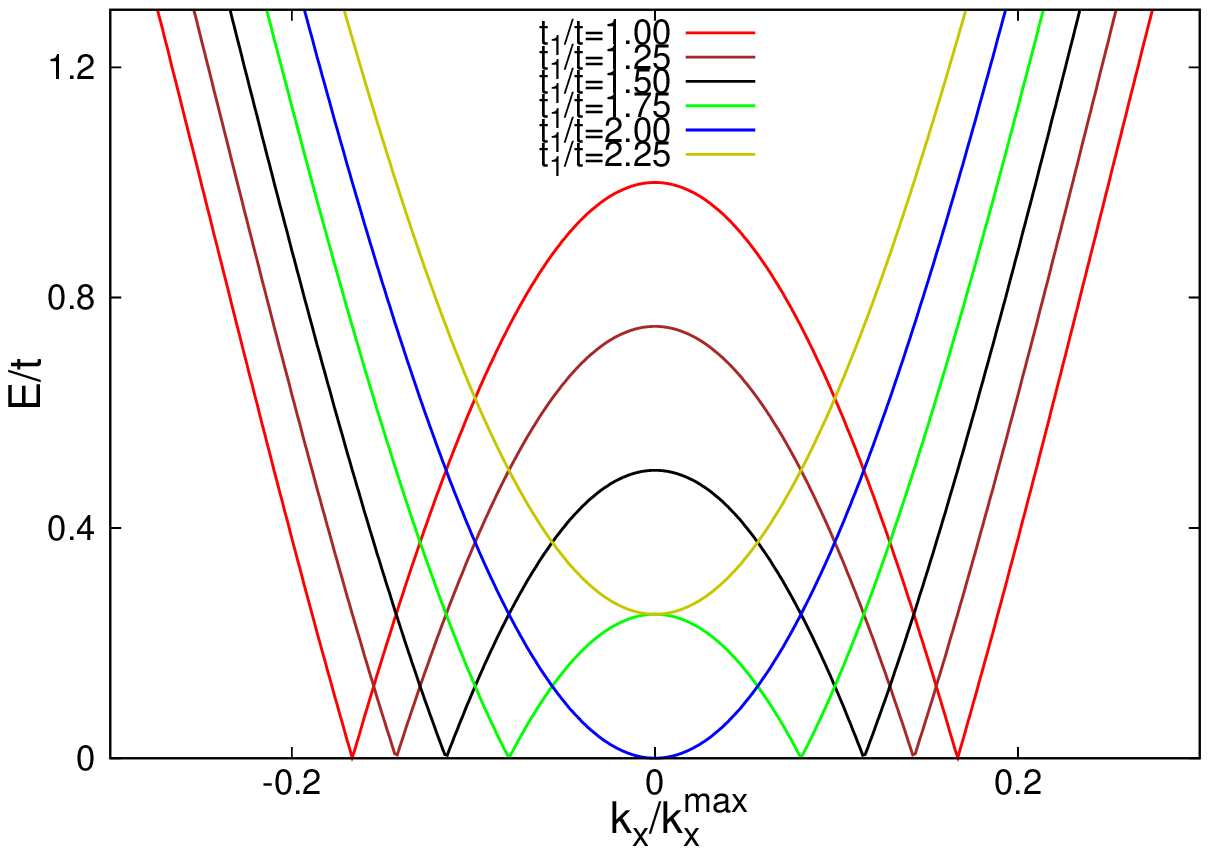}
\caption{\label{fig:merging}
%\col
%(Color online) 
Merging of two Dirac nodes with increasing anisotropy. For $t_1=2.25t$ there is a gap.
}
\end{center}
\end{figure}

The Hamiltonian of a tight-binding model with honeycomb structure (e.g., graphene) reads
in sublattice representation of the two-dimensional $\bk$ space \cite{castro09,abergel10}
\beq
H=\pmatrix{
0 &  -\sum_{j=1}^3t_je^{i{\bf a}_j\cdot\bk} \cr
-\sum_{j=1}^3t_je^{-i{\bf a}_j\cdot\bk} & 0 \cr
}
\label{ham1}
\eeq
with the basis vector of the sublattice with empty circles in Fig. \ref{fig:hexagon}
\beq
{\bf a}_1=a(0,-1), \ \ {\bf a}_{2,3}=\frac{a}{2}(\pm\sqrt{3},1)
\eeq
and with the lattice constant $a$ of this sublattice.
For fixed basis vectors ${\bf a}_j$ the positions of the two Dirac nodes depend 
on the hopping parameters $t_{1,2,3}$. Increasing one of them relative to the others
(i.e., breaking isotropy) moves the
Dirac nodes towards each other. This is accompanied by lifting the degenerate saddle
points of the spectrum. 
In particular, the Dirac nodes merge if two tunneling parameters are equal and the third one is
twice as large, for instance, for $t_1=2t_2=2t_3$ (cf. Fig. \ref{fig:merging}). For simplicity,
we consider subsequently only $t_2=t_3\equiv t$.
\begin{figure}[t]
\begin{center}
\includegraphics[width=8cm,height=7cm]{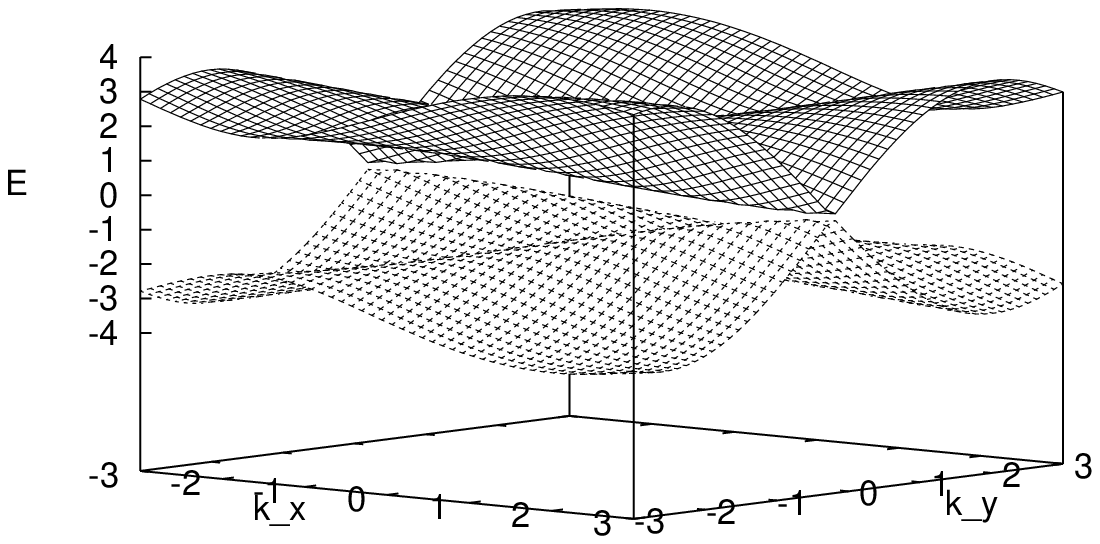}
\includegraphics[width=8cm,height=7cm]{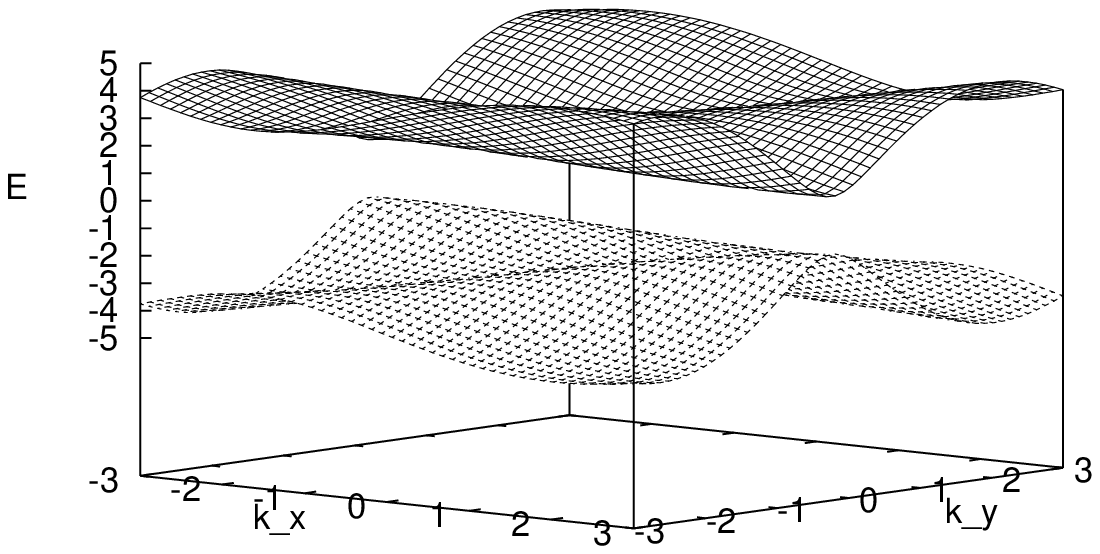}
\caption{\label{fig:open_bands}
%\col
%(Color online) 
Band structure of the honeycomb lattice for $t_1=2t$ (top) and $t_1=3t$ (bottom).
}
\end{center}
\end{figure}
At this point the Hamiltonian reads in the vicinity of the merged Dirac nodes
\beq
H=
\frac{p_x^2}{2m}\sigma_x+c_y p_y\sigma_y
\ ,
\label{hamilt_00}
\eeq
where $\sigma_j$ are Pauli matrices with the $2\times2$ unit matrix $\sigma_0$
%defined in Eq. (\ref{hamilt_00}) 
with coefficients
$m$ and $c_y$, which are related to the tunneling rates $t_j$ in a rather complex manner
\cite{montambaux09}. Thus, the tuning of $t_1$
allows us to measure the effect of the internode scattering on the transport properties. 

Adroguer et al. \cite{adroguer16} found with the Hamiltonian (\ref{hamilt_00}) for
the conductivity in the presence of a random scattering rate $\gamma$
\beq
\sigma_{xx}(E_F)\approx 0.197\ \frac{e^2\hbar}{\pi\gamma}\frac{2E_F}{m} \ ,
\ \ \ 
\sigma_{yy}(E_F)\approx 1.491\ \frac{e^2\hbar}{\pi\gamma}c_y^2
\ .
\label{cond00}
\eeq
The conductivity $\sigma_{yy}(E_F)$ diverges for vanishing disorder, while the behavior
of $\sigma_{xx}(E_F)$ diverges with $\gamma\sim0$ only for 
$E_F\ne 0$. On the other hand, the result is not unique in the limit $E_F\to0$, $\gamma\to0$. 

Since the conductivity is proportional to the diffusion coefficient of scattered electrons
at the Fermi energy $E_F$ due to the Einstein relation, the result in Eq. (\ref{cond00}) 
reflects also a strongly anisotropic diffusion coefficient at the anisotropic Dirac node
with $E_F=0$. This shall be studied in this paper in the absence of random scattering.
For potential opto-electronic applications it is important to analyze the optical conductivity.

\begin{figure}
\begin{center}
\includegraphics[width=9cm]{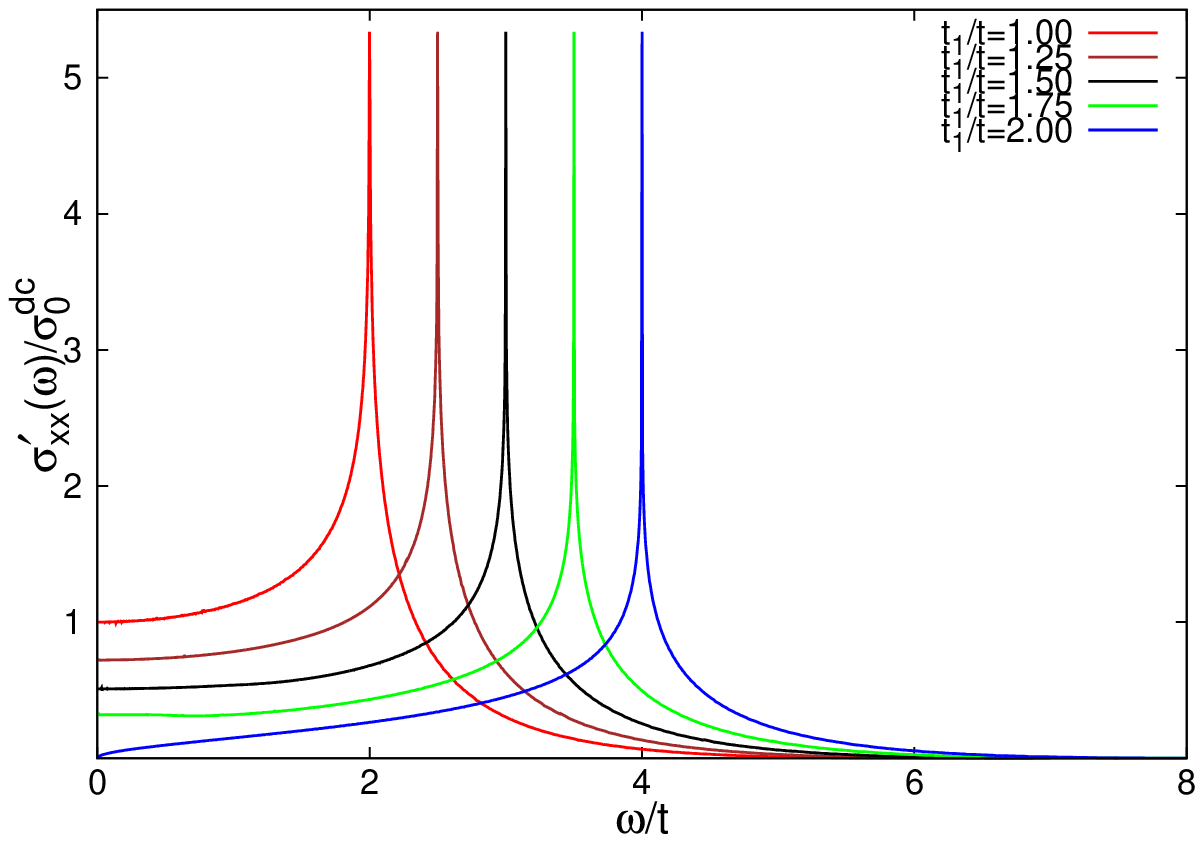}
\includegraphics[width=9cm]{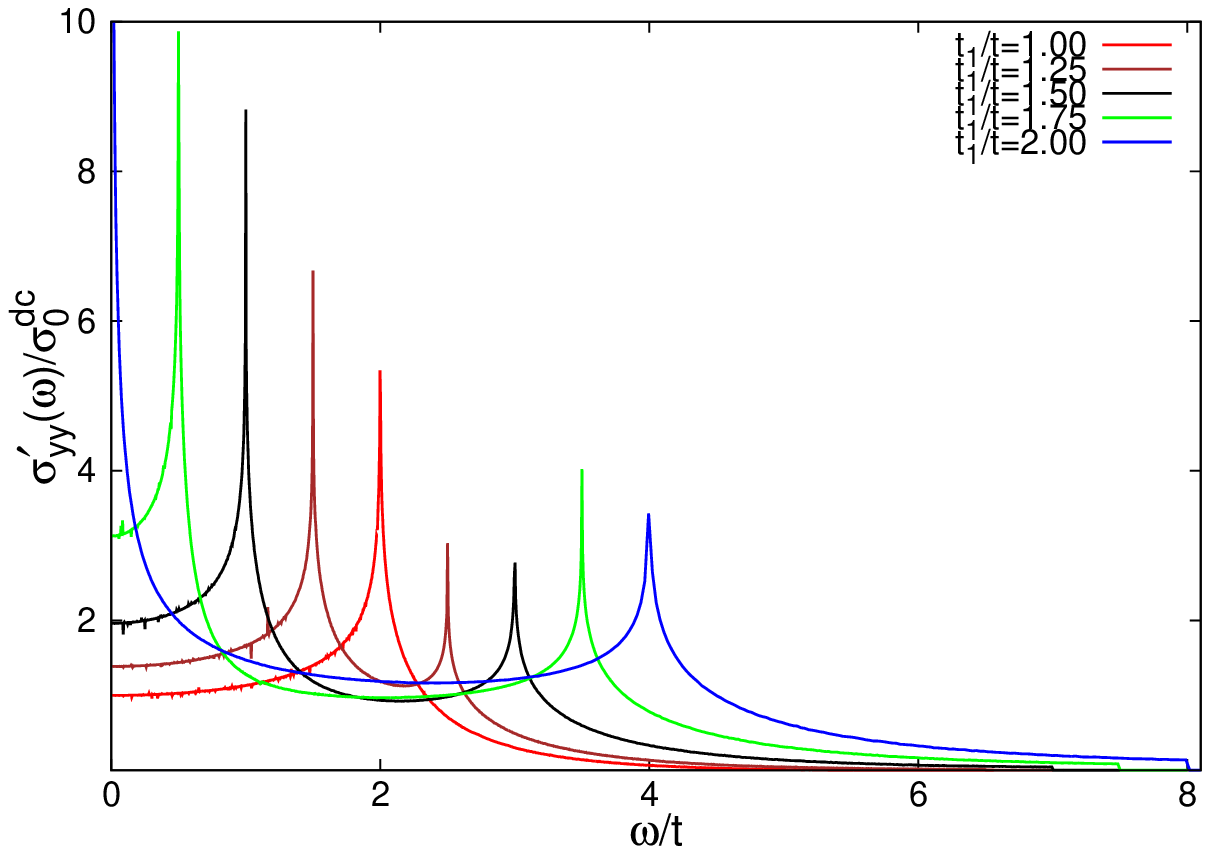}
\caption{\label{fig:opt_cond1}
%\col
%(Color online) 
Optical conductivities $\sigma_{xx}'$ and $\sigma_{yy}'$ with moving van Hove singularities
for different strengths of the anisotropy. The van Hove singularity for $\sigma_{yy}'$ of the 
isotropic system is split by the anisotropy into a pair of van Hove singularities, which move
apart with increasing anisotropy.
}
\end{center}
\end{figure}

\section{Optical conductivity}
\label{sect:opt_cond}

\begin{figure}
\begin{center}
\includegraphics[width=9cm]{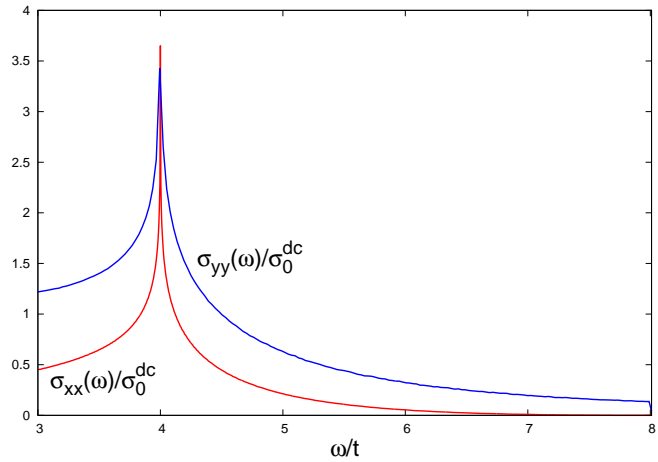}
\caption{\label{fig:opt_cond2}
%\col
%(Color online) 
High-frequency behavior of $\sigma_{xx}'$ and $\sigma_{yy}'$ near the van Hove singularity
at $\omega=4t$ of the anisotropic lattice with $t_1=2t$, where the effect of the anisotropy is weak.
}
\end{center}
\end{figure}

The linear response of the electronic system to an electromagnetic field of frequency $\omega$
is described by the Kubo formula of the optical conductivity. At temperature $T$ this reads
\beq
\sigma_{\mu\nu}(\omega)=-\frac{i}{\hbar}\lim_{\alpha\to0}\int_{BZ}\sum_{l,l'=0,1}\frac{\langle k,l|j_\mu|k,l'\rangle
\langle k,l'|j_\nu|k,l\rangle}{E_{k,l}-E_{k,l'}+\omega-i\alpha}
\frac{f_{\beta}(E_{k,l'})-f_{\beta}(E_{k,l})}{E_{k,l}-E_{k,l'}}\frac{d^2 k}{\Omega_{BZ}}
\label{kubo00}
\eeq
with the Fermi-Dirac distribution $f_\beta(E)=1/[1+\exp(\beta(E-E_F))]$, $\beta=1/k_b T$. $\Omega_{BZ}$
is the area of the Brillouin zone $BZ$, and $l,l'$ refers to the band index.
Its derivation can be found, for instance, in Ref. \cite{hill11}. This formula gives us for the Hamiltonian 
(\ref{hamilt_00}) with $T\sim0$, $E_F=0$ and $\omega>0$
\beq
Re [\sigma_{xx}(\omega)]\equiv\sigma_{xx}'\sim \omega^{1/2}I_x\ ,\ \ \ 
Re [\sigma_{yy}(\omega)]\equiv\sigma_{yy}'\sim \omega^{-1/2}I_y
\label{ocond11}
\eeq
for small $\omega$, where
$I_{x,y}$ are $\omega$-independent integrals. Thus, the anisotropy of the optical conductivity 
depends strongly on $\omega$. 
The corresponding optical conductivity of an isotropic Dirac node is independent of $\omega$.
For the full Hamiltonian (\ref{ham1}) and arbitrary values of $\omega$ the behavior of $\sigma_{xx}'$ 
and $\sigma_{yy}'$ is plotted in Figs. \ref{fig:opt_cond1} -- \ref{fig:opt_cond2}. For higher frequencies 
the behavior of the two conductivities is compared in Fig. \ref{fig:opt_cond2}.

The real part of the integrand in Eq. (\ref{kubo00}) contains a Dirac Delta function for $\alpha\sim0$,
which picks $ E_{k,l'}-E_{k,l}=\omega$. Thus $\sigma_{\mu\mu}'=0$ for $\omega$ smaller than the gap
\cite{hill11}.

\section{Anomalous diffusion}
\label{sect:diffusion}

The motion of particles (e.g., electrons) is characterized by the mean square displacement of a particle position,
a concept which has been applied to classical as well as to quantum systems \cite{thouless74}. It provides our basic 
understanding for a large number of transport phenomena, such as the metallic behavior in electronic systems.
Starting point is the transition probability for a particle, governed by the Hamiltonian $H$,
to move from the site $\br'$ on a lattice to another lattice site $\br$ within the time $t$: 
\beq
P_{\br\br'}(t)=\sum_{j,j'}
%\langle|
|\langle \br,j|e^{-iHt}|\br',j'\rangle|^2
%\rangle_d
\ .
\label{trans_prob}
\eeq
Here the indices $j,j'$ refer to different bands of the system, represented in the Hamiltonian 
(\ref{ham1}) or (\ref{hamilt_00}) by Pauli matrices.
With the expression (\ref{trans_prob}) we obtain, for instance, the mean square displacement as
\[
\langle (r_\mu-r'_\mu)^2\rangle= \sum_r (r_\mu-r_\mu')^2 P_{\br\br'}(t)
\ \ \ (\mu=x,y)
\ .
\]
After integration with respect to time this becomes according to Eq. (\ref{time-av})
\beq
\eps^2\sum_\br (r_\mu-r_\mu')^2 \int_0^\infty P_{\br\br'}(t) e^{-\eps t}dt
=\int D_\mu(\eps,E) dE
\eeq
with $D_\mu(\eps,E)$ defined in Eq . (\ref{diff_01}).
%There is an additional prefactor $\eps$ in comparison with the expression (\ref{time-av}) 
%to get a finite expression in the case of diffusion.
Thus, we obtain the integral of the diffusion coefficient $D_\mu(\eps,E)$ for 
a particle of energy $E$ with respect to all energies. In the case of a Fermi gas only fermions
at the Fermi energy contribute at sufficiently low temperatures. Therefore,
we study this coefficient for particles at a fixed Fermi energy $E$. Next, 
$D_\mu(\eps,E)$ is calculated for a node with linear and/or parabolic spectrum, as given 
for the merging point of two Dirac nodes in Eq. (\ref{hamilt_00}).
With the expression (\ref{diff_1}) for the diffusion coefficient we obtain from 
Eqs. (\ref{diff_x}) and (\ref{diff_y}) 
\beq
D_x\approx %5.32
{\bar D}_x \eps^{1/2}\ ,
\ \ \ 
D_y\approx %1.94\ 
{\bar D}_y \eps^{-1/2}
\ ,
\label{diff_coeff}
\eeq
where the $\eps$--independent coefficients ${\bar D}_x$ and ${\bar D}_y$ are integrals
given in App. \ref{sect:anisotropic}.

These expressions for the diffusion coefficients reflect the result of a divergent $\sigma_{yy}$ for 
$\gamma\to 0$ in Eq. (\ref{cond00}) and clarifies the transport behavior at $E_F=0$ and $\gamma=0$,
previously found in Ref. \cite{adroguer16}. It corresponds with the asymptotic time ($\tau$) behavior
of superdiffusion ($\sim \tau^{3/2}$) in $y$ direction and subdiffusion (i.e., $\sim \tau^{1/2}$) 
in $x$ direction. For an isotropic node
we get $D^{(1)}\sim 4\pi/3$ for a linear (Dirac) dispersion and $D^{(2)}\sim 8\pi/3$ 
for quadratic dispersion, respectively (cf. App. \ref{sect:isotropic}).

We anticipate that additional disorder scattering would replace $\eps$ by disorder strength $\gamma$
which reduces the superdiffusive behavior to normal diffusion in the $y$ direction, whereas the subdiffusive 
behavior in the $x$ direction would persist. This agrees with the behavior of the conductivity in Eq. (\ref{cond00}). 

\begin{figure}
\begin{center}
\includegraphics[width=9cm]{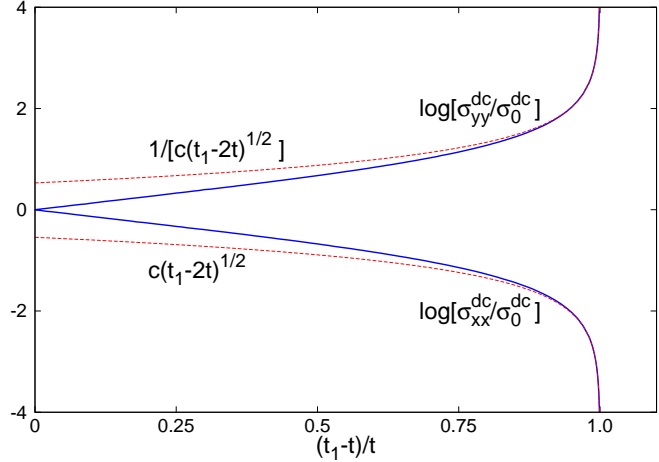}
\caption{\label{fig:dc_cond}
%\col
%(Color online) 
Scaling behavior of the conductivity for $\omega\sim0$ with the strength of the anisotropy.
}
\end{center}
\end{figure}

\section{Discussion}
\label{sect:discussion}

Transport properties 
are very sensitive to the topology of the band structure. We have studied this in terms of
the optical conductivity and mean-square displacement for a system, where the topology of 
the band structure is characterized by two bands with two nodes and saddle points between these nodes
(i.e., a vanishing gradient of the spectrum $\nabla_\bk E_\bk=0$ with 
different curvatures in different directions).
For our study it is essential that in real space we have a lattice with 
global symmetries. For instance, there is a sublattice symmetry due to the two 
degenerate triangular 
lattices forming a honeycomb lattice and a discrete threefold isotropy. Then the band structure
of the infinite lattice
is a continuous compact manifold. Continuous deformations of the latter can be achieved by 
breaking global lattice symmetries. For example, breaking the sublattice symmetry opens
the Dirac nodes and creates a gap between the two bands \cite{castro09,abergel10}. The
size of the gap increases continuously with increasing sublattice asymmetry.
Breaking the global isotropy of the lattice by changing the electronic hopping rate in one direction,
breaks the isotropy of the nodal structure and moves the nodes to different locations in $\bk$ 
space \cite{montambaux09,montambaux09a,delplace10,adroguer16}. At the same time the 
degeneracy of the saddle points is lifted. The reduction of the saddle point value upon an
increasing anisotropy is visualized in Fig. \ref{fig:merging}.

%-- transport properties:

Spectral properties are difficult to observe directly; it is easier to
analyze them indirectly through transport measurements. For instance, the saddle points
in the spectrum
lead to van Hove singularities in the optical conductivity (cf. Figs. \ref{fig:opt_cond1} 
and \ref{fig:opt_cond2}).
Nodes, on the other hand, are characterized by a kind of universal transport
behavior when random scattering is suppressed, as discussed in Sect. \ref{sect:introduction}. This limit reveals details of the
spectrum at the nodes. More directly, though, is the measurement of the optical
conductivity for low frequencies.

The main effect of a globally broken isotropy on the optical conductivity $\sigma_{yy}'$
is the appearance of two van Hove singularities rather than one, 
in contrast to the isotropic case. This is visualized in Fig. \ref{fig:opt_cond1}:
For the isotropic case $t_1=t$ of the Hamiltonian (\ref{ham1}) the saddle points are degenerate.
The conductivities (red curves in Fig. \ref{fig:opt_cond1}) indicate a single van Hove singularity at 
$\omega=2t$. In the anisotropic case $t_1>t$, on the other hand, the degeneracy of the saddle 
points is lifted to one van Hove singularity at $\omega <2t$ and another one at $\omega>2t$. 
In the special case of $t_1=2t$ one saddle point appears at $\omega=0$ and the other 
one at $\omega=4t$. 

The frequency $\omega$ of the external electromagnetic field creates a finite length scale (wave length) 
for the electronic system
through the relation $\lambda=v_F/\omega$. The Fermi velocity $v_F$ is the spectral slope near the
Dirac node. 
This length scale is similar to the mean-free 
path $l_m=v_F\tau$, created by random scattering with scattering time $\tau$. In our 
calculation of Sect. \ref{sect:diffusion} the role of $\tau$ is played by $\hbar/\epsilon$.
For the optical conductivity the electronic wave length $\lambda$ sets a finite scale in the 
(graphene) sample. The relevance of these two length scales is reflected by the similarity
of the optical conductivity in Eq. (\ref{ocond11}) and the diffusion coefficients in Eq. (\ref{diff_coeff})
in terms of $\omega$ and $\epsilon$, respectively. An increasing length creates
an increasing $\sigma_{yy}$ ($D_y$) and a decreasing $\sigma_{xx}$ ($D_x$).

In isotropic graphene the hopping parameter is $t=2.8 {\rm eV}$ and the Fermi velocity is 
$v_F=$10$^6$m/s. If we choose for the frequency of the external electromagnetic field 
$\omega=2t/\hbar\approx 8\cdot 10^{15}~ {\rm s}^{-1}$ (extreme ultraviolet), we have for 
the conductivity at the van Hove singularity an electronic wave length of 
$\lambda= v_F/\omega\approx 2.3\cdot 10^{-10}$m. Through a deformation of the lattice
we can tune the frequency of the van Hove singularity between 0 and $4t$, which would allow us 
to detect this singularity with the optical conductivity over a wide range of frequencies.

For very low frequencies (i.e., for very large length scales) the conductivity satisfies a scaling
behavior with respect to anisotropy parameter $\Delta=(t_1-t)/t$. There is a critical 
point $\Delta=1$, which is approached by the conductivity with a power law, as visualized
in Fig. \ref{fig:dc_cond}. For every particular value of $\Delta$ both conductivities obey a 
beautiful phenomenological constrain condition
\begin{equation}
\label{eq:Constraint}
\sqrt{\sigma^{dc}_{xx}\sigma^{dc}_{yy}} = \sigma^{dc}_0. 
\end{equation}
Microscopically, Eq.~(\ref{eq:Constraint}) is a consequence of the quantum diffusion current conservation.

An important question is how the isotropy can be globally broken in a honeycomb
lattice. In graphene, for instance, we can apply uniaxial pressure or pull the material
in one direction. It seems to be unrealistic, though, that one could reach the point of degenerate
nodes (i.e., $t_1=2t$) by this method. Another possibility is to use ``artificial'' graphene \cite{haldane08}
in the form of a photonic crystal \cite{rechtsman13,keil13} or a microwave metamaterial \cite{cheng16}, which 
can be designed in the laboratory with any kind of lattice structure. A disadvantage is that the particles
are photons rather than electrons, for which the conductivity cannot be measured directly. 
A third option is spontaneous breaking of the isotropy through electron-phonon interaction. Analogous
to the dimerization in 1D materials \cite{heeger88}, dimerization can also occur in 2D materials,
for instance, in the form of Kekul\'e order \cite{gutierrez16}. Our case
of $t_1>t$ would represent a similar order.

\section{Conclusions}

The spectral properties near a node in a two-band system determines the electronic transport. While an isotropic
node with linear spectrum (Dirac node) creates isotropic diffusion, an anisotropic node with a linear spectrum in
one direction and a quadratic spectrum in the other direction leads to anisotropic transport with subdiffusive
in one direction and superdiffusive behavior in the other direction, respectively. The fact that a lattice system
is associated with a compact manifold of the band structure opens up the possibility to study topological transitions
of compact manifolds by tuning the global lattice symmetry, such as sublattice symmetries or isotropy. This is
particularly promising for photonic metamaterials, where the lattice structure is easy to modify \cite{cheng16}.

\vskip0.3cm

Acknowledgment: We are grateful to Gilles Montambaux for discussing his study of merging Dirac nodes.
This work was supported by a grant of the Julian Schwinger Foundation.

\appendix

\section{Transition probability and mean square displacement
%(from 1412)
}
\label{sect:trans_prob}

Averaging over large times
\beq
\eps\int_0^\infty\langle (r_\mu-r'_\mu)^2\rangle  e^{-\eps t}dt
%\lim_{\eps\to 0}
=\eps\sum_\br (r_\mu-r_\mu')^2 \int_0^\infty P_{\br\br'}(t) e^{-\eps t}dt
\label{time-av}
\eeq
provides the asymptotic behavior 
%\eps^2\int_0^\infty \langle (r_\mu-r'_\mu)^2\rangle  e^{-\eps t}dt
$\sim A \eps^{-\alpha}$ for this expression if the long-time behavior of the mean square displacement is
\beq
\langle (r_\mu-r'_\mu)^2\rangle\sim A t^\alpha
\ .
\eeq
Special cases are diffusion for $\alpha=1$ and ballistic transport for $\alpha=2$.

The time integral in Eq. (\ref{time-av}) can also be written in terms of the Green's function 
$G(z)=(H-z)^{-1}$ as an energy integral
\beq
\int_0^\infty P_{\br\br'}(t) e^{-\eps t}dt
=\frac{1}{\pi}\int Tr_2\left\{G_{\br\br'}(E-i\eps)\left[
G_{\br'\br}(E+i\eps) -G_{\br'\br}(E-i\eps)\right]\right\}dE
\label{msd1}
\ ,
\eeq
where $Tr_2$ is the trace with respect to the spinor index. The one-particle Green's function
is defined as the resolvent $G(z)=(H-z)^{-1}$ of the Hamiltonian $H$, and $G_{\br0}(E+i\eps)$ describes
the propagation of a particle with energy $E$ from the origin to a site $\br$.
Then the expression in Eq. (\ref{time-av}) becomes
\beq
\eps^2\sum_\br (r_\mu-r_\mu')^2 \int_0^\infty P_{\br\br'}(t) e^{-\eps t}dt
=\int D_\mu(\eps,E) dE
\eeq
with
\beq
D_\mu(\eps,E)=\frac{1}{\pi}\eps^2\sum_\br (r_\mu-r_\mu')^2
Tr_2\left\{G_{\br\br'}(E-i\eps)\left[ G_{\br'\br}(E+i\eps) -G_{\br'\br}(E-i\eps)\right]\right\}
\ .
\label{diff_01}
\eeq
There is an additional prefactor $\eps$ in comparison with the expression (\ref{time-av}) to get a finite
expression in the case of diffusion.
Thus, we obtain the integral of the diffusion coefficient $D_\mu(\eps,E)$ for 
a particle of energy $E$ with respect to all energies.
Subsequently, we study this coefficient for particles at a fixed energy $E$. 
The coefficient (\ref{diff_01}) for a two-band system of 
non-interacting particles with Hamiltonian $H_\bk$ at the node with energy $E=0$ is proportional to
\[
D_\mu(\eps,0)=-\eps^2\int_\bk Tr_2\left(\left[\frac{\partial^2}{\partial k_\mu^2}
(H_\bk-i\eps)^{-1}\right]\left[(H_\bk+i\eps)^{-1}-(H_\bk-i\eps)^{-1}\right]\right)
\ ,
\]
where $\int_\bk$ is the integral with respect to the Brillouin zone of the lattice,
\[
=-2\eps^2\int_\bk Tr_2\left[(H_\bk^2+\eps^2)^{-1}
\frac{\partial H_\bk}{\partial k_\mu}(H_\bk-i\eps)^{-1}\frac{\partial H_\bk}{\partial k_\mu}(H_\bk-i\eps)^{-1}\right]
\]
\beq
+\eps^2\int_\bk Tr_2\left[(H_\bk^2+\eps^2)^{-1}
\frac{\partial^2 H_\bk}{\partial k_\mu^2}(H_\bk-i\eps)^{-1}\right]
-\eps^2\int_\bk Tr_2\left[(H_\bk-i\eps)^{-2}
\frac{\partial H_\bk}{\partial k_\mu}(H_\bk-i\eps)^{-2}\frac{\partial H_\bk}{\partial k_\mu}\right]
\ .
\label{diff_1}
\eeq

\section{Diffusion at an isotropic node}
\label{sect:isotropic}

\beq
H_\bk^{(1)} = k_1\sigma_1+k_2\sigma_2 , \ \ \ H_\bk^{(2)} = (k_1^2-k_2^2)\sigma_1+2k_1k_2\sigma_2
\ .
\eeq
Using the expression (\ref{diff_1}), a straightforward calculation yields
\beq
D_\mu^{(1)}=4\pi\eps^2\int_0^\lambda\frac{\eps^4+4\eps^2k^2-k^4}{(k^2+\eps^2)^4}kdk
=4\pi\int_0^{\lambda/\eps}\frac{1+4\kappa^2-\kappa^4}{(\kappa^2+1)^4}\kappa d\kappa
\sim \frac{4\pi}{3} \ \ \  (\kappa_j=k_j/\eps)
\eeq
\beq
D_\mu^{(2)}=\pi\eps^{2}\int_0^\lambda\frac{k^2(16\eps^4+64\eps^2k^4-16k^8)}{(k^4+\eps^2)^4}kdk
=16\pi\int_0^{\lambda/\sqrt{\eps}}\frac{\kappa^2(1+4\kappa^4-\kappa^8)}{(\kappa^4+1)^4}\kappa d\kappa
\sim\frac{8\pi}{3}  \ \ \  (\kappa_j=k_j/\sqrt{\eps})
\ ,
\eeq
where the numerical values of the integral are obtained for $\eps\sim 0$.
Thus, both coefficients are finite for $\eps\to0$ and describe diffusion.

\section{Diffusion at an anisotropic node}
\label{sect:anisotropic}

From the expression (\ref{diff_1}) we obtain
\beq
D_x=2\eps^{2}\int_\bk\frac{k_1^2[5\eps^4+\eps^2(26k_1^4+10k_2^2)-11k_1^8-6k_1^4k_2^2+5k_2^4]}{(k_1^4+k_2^2+\eps^2)^4}
%=8\int_\kappa\frac{\kappa^2}{(\kappa^4+1)^3} \ \ \  (\kappa_j=k_j/\sqrt{\eps})
\eeq
\beq
D_y=-4\eps^{2}\int_\bk\frac{\eps^4+\eps^2(2k_1^4-12k_2^2)+k_1^8+4k_1^4k_2^2+3k_2^4}{(k_1^4+k_2^2+\eps^2)^4}
%=8\int_\kappa\frac{\kappa^2}{(\kappa^4+1)^3} \ \ \  (\kappa_j=k_j/\sqrt{\eps})
\ .
\eeq
After the rescaling $k_1\to\kappa_1=k_1/\sqrt{\eps}$ and $k_2\to\kappa_2=k_2/\eps$ these expressions become
\beq
D_x=2\eps^{1/2}\int_\kappa\frac{\kappa_1^2[5+26\kappa_1^4+10\kappa_2^2
-11\kappa_1^8-6\kappa_1^4\kappa_2^2+5\kappa_2^4]}{(\kappa_1^4+\kappa_2^2+1)^4}
\label{diff_x}
\eeq
\beq
D_y=-4\eps^{-1/2}\int_\kappa\frac{1+2\kappa_1^4-12\kappa_2^2+\kappa_1^8+4\kappa_1^4\kappa_2^2+3\kappa_2^4}{(\kappa_1^4+\kappa_2^2+1)^4}
\ .
\label{diff_y}
\eeq


\begin{thebibliography}{99}

\bibitem{novoselov05}K.S. Novoselov, A.K. Geim, S.V. Morozov, D.
Jiang, M.I. Katsnelson, I.V. Grigorieva, S.V. Dubonos, and A.A.
Firsov, Nature {\bf 438}, 197 (2005).

\bibitem{castro09}
A.H. Castro Neto, F. Guinea, N.M.R. Peres, K.S. Novoselov,  
and A.K. Geim, Rev. Mod. Phys. {\bf 81} 109 (2009).

\bibitem{abergel10}
D.S.L. Abergel, V. Apalkov, J. Berashevich, K. Ziegler and T. Chakraborty, 
%'Properties of graphene: a theoretical perspective', 
Advances in Physics {\bf 59}, 261 (2010).

\bibitem{watanabe04}
K. Watanabe, T. Taniguchi, and H. Kanda, Nature Materials {\bf 3}, 404 - 409 (2004). 

\bibitem{thouless74}
D.J. Thouless, Phys. Rep. {\bf 13}, 93 (1974).

\bibitem{chen08}
J.-H. Chen, C. Jang, S. Adam, M.S. Fuhrer, E.D. Williams and M. Ishigami,
Nature Physics {\bf 4}, 377 - 381 (2008). 

\bibitem{chen09}
J.-H. Chen, W.G. Cullen, C. Jang, M.S. Fuhrer, and E.D. Williams,
Phys. Rev. Lett. {\bf 102}, 236805 (2009).

\bibitem{montambaux09}
G. Montambaux, F. Pi\'echon, J.-N. Fuchs, and M.O. Goerbig, Phys. Rev. B {\bf 80}, 153412 (2009).

\bibitem{montambaux09a}
G. Montambaux, F. Pi\'echon, J.-N. Fuchs, and M.O. Goerbig, Eur. Phys. J B {\bf 72}, 509 (2009).

\bibitem{delplace10}
P. Delplace and G. Montambaux, Phys. Rev. B {\bf 82}, 035438 (2010).

\bibitem{adroguer16}
P. Adroguer, D. Carpentier, G. Montambaux, and E. Orignac,
Phys. Rev. B {\bf 93}, 125113 (2016).

\bibitem{faye14}
J.P.L. Faye, S.R. Hassan, D. S\'en\'echal, Phys. Rev. B {\bf 89}, 115130 (2014).

\bibitem{ziegler11a}
K. Ziegler, E. Kogan, E. Majernikova and S. Shpyrko,
Phys. Rev. B {\bf 84}, 073407 (2011).

\bibitem{ziegler11b}
K. Ziegler and E. Kogan, EPL {\bf 95}, 36003 (2011). 

\bibitem{hill11}
A. Hill, A. Sinner and K. Ziegler, New J. Phys. {\bf 13}, 035023 (2011).

\bibitem{haldane08}
F.D.M. Haldane and S. Raghu, Phys. Rev. Lett. {\bf 100}, 013904 (2008); 
S. Raghu and F.D.M. Haldane, Phys. Rev. A {\bf 78}, 033834 (2008).

\bibitem{rechtsman13}
M.C. Rechtsman, J.M. Zeuner, Y. Plotnik, Y. Lumer, D. Podolsky,
F. Dreisow, S. Nolte, M. Segev, and A. Szameit, Nature {\bf 496}, 196 (2013) %, ISSN 0028-0836, letter

\bibitem{keil13}
Keil R. et al. % The random mass Dirac model and long-range correlations on an integrated optical platform. 
Nat. Commun. {\bf 4}:1368 %doi: 10.1038/ncomms2384 (2013).
%R. Keil, J.M. Zeuner, F. Dreisow, M. Heinrich, A. Tunnermann, S. Nolte \& A. Szameit,

\bibitem{cheng16}
X. Cheng, C. Jouvaud, X. Ni,	S. Hossein Mousavi,	A.Z. Genack	\& A.B. Khanikaev,
Nature Mater. {\bf 15}, 542-548 (2016).

\bibitem{heeger88}
A.J. Heeger, S. Kivelson, J.R. Schrieffer, and W.P. Su, Rev. Mod. Phys. {\bf 60}, 781 (1988).

\bibitem{gutierrez16}
Ch. Gutierrez et al., Nature Phys. {\bf 12}, 950–958 (2016).

\end{thebibliography}
\end{document}